# Peculiar magnetic and magneto-transport properties in a non-centrosymmetric self-intercalated van der Waals ferromagnet Cr$_5$Te$_8$

Banik Rai[1], Sandip Kumar Kuila[2], Rana Saha[3,4], Sankalpa Hazra[5], Chandan De[6,7], Jyotirmoy Sau[1], Venkatraman Gopalan[5,6], Partha Pratim Jana[2], Stuart S. P. Parkin[3], Nitesh Kumar[1*]

[1]Department of Condensed Matter and Materials Physics, S. N. Bose National Centre for Basic Sciences, Salt Lake City, Kolkata 700106, India

[2]Department of Chemistry, Indian Institute of Technology, Kharagpur 721302, India

[3]Max Planck Institute of Microstructure Physics, Weinberg 2, Halle (Saale) 06120, Germany

[4]Department of Chemistry, Indian Institute of Science Education and Research, Yerpedu, Tirupati 517619, India

[5]Department of Materials Science and Engineering, The Pennsylvania State University, University Park, PA 16802, USA

[6]Department of Physics, The Pennsylvania State University, University Park, PA 16802, USA

[7]2D Crystal Consortium, Materials Research Institute, The Pennsylvania State University, University Park, PA, USA

**Abstract**

Trigonal Cr$_5$Te$_8$, a self-intercalated van der Waals ferromagnet with an out-of-plane magnetic anisotropy, has long been known to crystallize in a centrosymmetric structure. However, optical second harmonic generation experiments, together with comprehensive structural analysis, indicate that this compound rather adopts a non-centrosymmetric structure. Lorentz transmission electron microscopy reveals the presence of Néel-type skyrmions, consistent with its non-centrosymmetric structure. A large anomalous Hall conductivity of 102 $\mathbf{\Omega^{-1} cm^{-1}}$ at low temperature stems from intrinsic origin, which is larger than any previously reported values in the bulk Cr-Te system. Notably, spontaneous topological Hall resistivity arising from the skyrmionic phase has been observed. Our findings not only elucidate the unique magnetic and magneto-transport properties of non-centrosymmetric trigonal Cr$_5$Te$_8$, but also open new avenues for investigating the effects of broken inversion symmetry on material properties and their potential applications.

**Introduction**

Van der Waals (vdW) magnets have recently garnered considerable attention due to their potential technological applications.[1–3] Magnetic anisotropy can stabilize long-range magnetic ordering in such systems even in the monolayer limit.[1,2] Among vdW compounds, transition metal dichalcogenides (TMDs) are particularly noteworthy due to their diverse physical properties, spanning from narrow bandgap semiconductors[4,5] to topological semimetals[6,7] and superconductors.[8] These systems also offer flexibility for controlled intercalation within the vdW gaps, leading to phenomena such as superconductivity,[9,10] chirality-induced topological Hall effect,[11] intrinsic anomalous Hall effect,[12,13] etc. CrTe$_2$, in particular, has attracted particular attention because it is a rare example of vdW ferromagnetic TMD.[14–17] In the ultrathin limit, the magnetic easy axis of CrTe$_2$ is out-of-plane, which becomes in-plane as the thickness increases.[16] By intercalating additional Cr atoms into the vdW gaps, an out-of-plane magnetic easy axis can be achieved in bulk samples.[18] Cr$_5$Te$_8$ is one such self-intercalated compound that is known to exist in two distinct phases viz, trigonal (tr) Cr$_5$Te$_8$ and monoclinic (m) Cr$_5$Te$_8$.[19] These phases differ slightly in composition with the monoclinic phase being stable in the range of 59.6-61.5 at% Te, while the trigonal phase is stable in the range of 61.8-62.5 at% Te.[19] The Curie temperature of Cr$_5$Te$_8$ ranges from 180 K to 230 K and is highly susceptible to the amount of Cr intercalation. According to the available literature, tr-Cr$_5$Te$_8$ and m-Cr$_5$Te$_8$ crystallize in space groups $P\bar{3}m1$ and C2/m respectively,[19] both being centrosymmetric. In particular, topological Hall effect which stems from the scalar spin chirality (SSC),[20,21] has been observed in tr-Cr$_5$Te$_8$.[22] SSC is associated with a non-coplanar spatial arrangement of spins $\left(\chi_{ijk} = \boldsymbol{S}_i \cdot (\boldsymbol{S}_j \times \boldsymbol{S}_k)\right)$, often stabilized by the competition between the symmetric Heisenberg interaction (HI) and the antisymmetric Dzyaloshinskii-Moriya interaction (DMI), the latter being absent in centrosymmetric systems.[23,24] Furthermore, a DMI-induced skyrmionic phase has been observed in Cr-rich Cr$_{1.3}$Te$_2$, where the crystal structure was found to be non-centrosymmetric.[25] This discrepancy underscores contradictory claims regarding the structure and thereby their corresponding magnetic and electronic properties.

In this work, we have investigated the crystal structure, magnetic, and transport properties of tr-Cr$_5$Te$_8$. Through detailed single crystal and powder X-ray diffraction (SCXRD and PXRD), together with second harmonic generation (SHG) experiments, we show that the compound crystallizes in a non-centrosymmetric space group. Alongside the anomalous Hall effect, we observed a significant spontaneous topological Hall effect, originating from the Néel-type skyrmion phase in tr-Cr$_5$Te$_8$. The presence of the Néel skyrmions was directly observed using Lorentz transmission electron microscopy (LTEM).

---

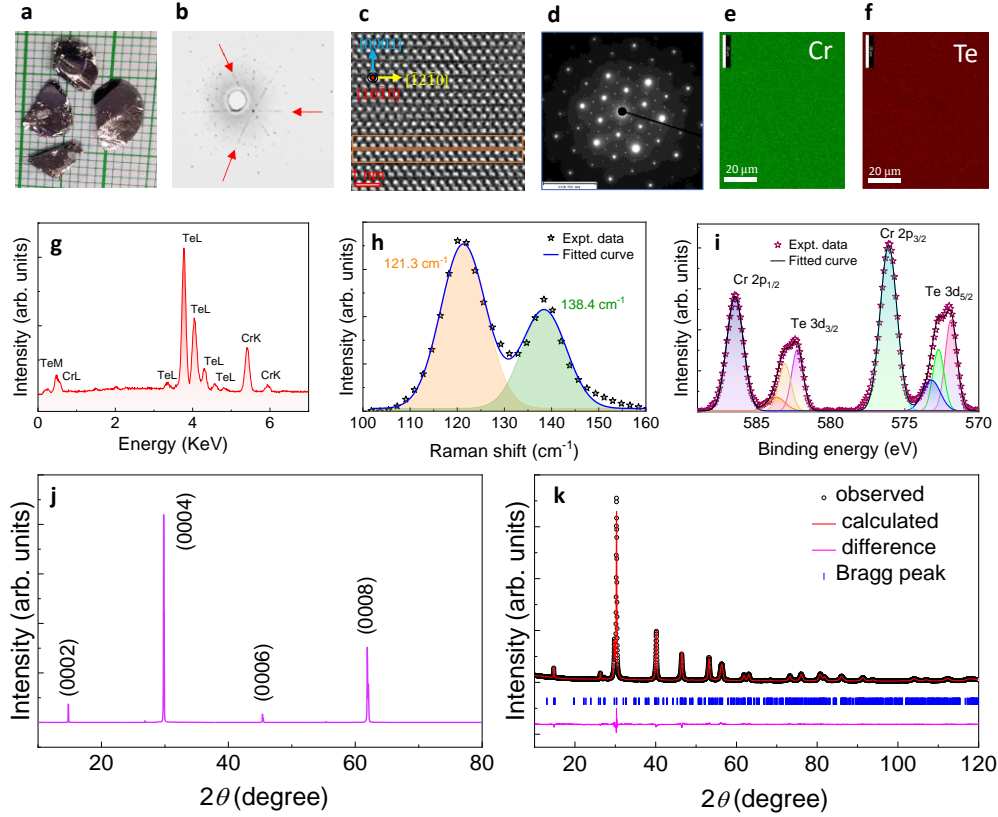

**Figure 1. Structural characterization of tr-$Cr_5Te_8$.** (a) Image of as-grown single crystals of tr-$Cr_5Te_8$. (b) Laue diffraction pattern of tr-$Cr_5Te_8$ single crystal along [0001] direction. Red arrows show the three-fold symmetric points. (c) HRTEM image of tr-$Cr_5Te_8$ viewed along [10$\bar{1}$0] direction. The Cr-Te-Cr triple layers are highlighted with the saffron rectangles. (d) SAED pattern of tr-$Cr_5Te_8$. (e)-(f) EDXS mapping of elements of tr-$Cr_5Te_8$. (g) A typical EDXS spectra of elemental composition of tr-$Cr_5Te_8$. (h) Raman spectra of as grown tr-$Cr_5Te_8$ single crystal. (i) Resolved XPS spectra of as grown tr-$Cr_5Te_8$ single crystal. (j) XRD pattern of tr-$Cr_5Te_8$ single crystal obtained by exposing the plane crystal facet to the X-ray beam. (k) Rietveld refinement of powder X-ray diffraction (PXRD) data of tr-$Cr_5Te_8$ refined with $P\bar{3}m1$ space group. The refinement parameters are as follow: $R(obs)$ = 2.51, $wR(obs)$ = 3.61, $R(all)$ = 3.12, $wR(all)$ = 3.73, GOF = 1.96, $Rp$ = 3.96, $wRp$ = 5.18.

## 1. Results and Discussions

Figure 1a shows the as-grown single crystals of tr-$Cr_5Te_8$, each measuring several millimeters in size. The Laue diffraction pattern captured in the reflected mode along the $c$-axis of the crystal is shown in Figure 1b. The pattern consists of three-fold symmetric points (highlighted by red arrows), confirming the trigonal structure of tr-$Cr_5Te_8$. The high-resolution transmission electron microscopy (HRTEM) image depicted in Figure 1c, taken along the [10$\bar{1}$0] direction, illustrates the layered structure of tr-$Cr_5Te_8$ along the [0001] direction. The selected area electron diffraction (SAED) pattern, shown in Figure 1d, features hexagonally arranged diffraction spots, consistent with the trigonal structure of tr-$Cr_5Te_8$. The energy dispersive X-ray spectroscopy (EDXS) elemental mapping of Cr (Figure 1e) and Te (Figure 1f) demonstrates the uniform distribution of elements throughout the grown crystal. Figure 1g shows the typical EDXS spectra of tr-$Cr_5Te_8$. The elemental composition as checked by EDXS (see Table S1) is 38.06 at% Cr and 61.94 at% Te, which is within the homogeneity range of tr-$Cr_5Te_8$.[19] The Raman spectra shown in Figure 1h shows two characteristic peaks at 121.3 cm$^{-1}$ and 138.4 cm$^{-1}$ corresponding to the out-of-plane and in-plane modes of the phonon vibration of tr-$Cr_5Te_8$. Figure 1i presents the resolved X-ray photoemission spectra (XPS) of tr-$Cr_5Te_8$. The sharp peaks at approximately 586 eV and 576 eV correspond to the Cr $2p_{1/2}$ and $2p_{3/2}$ core levels, respectively. In contrast, the Te $3d_{3/2}$ and $3d_{5/2}$ peaks at around 583 eV and 572 eV exhibit significant distortion, suggesting the presence of overlapping peaks. A detailed peak fitting analysis enables the deconvolution of the Te peaks into three distinct components. Structural analysis based on X-ray diffraction (discussed later) also suggests the existence of three different types of Te atoms (see Table S5) based on their coordination with Cr atoms. Given that XPS is sensitive to local chemical environments, the differences in the coordination of Te atoms likely contribute to the splitting of the Te peak into multiple components, resulting in the observed distortion of the overall peak profile. Figure 1j shows the XRD pattern obtained by exposing one of the flat-surfaces of a plate-like single crystal to the X-ray beam. The pattern only consists of (000$l$) peaks indicating that the plane exposed is the *ab*-plane. The absence of other peaks suggests a good crystallinity of the crystal.

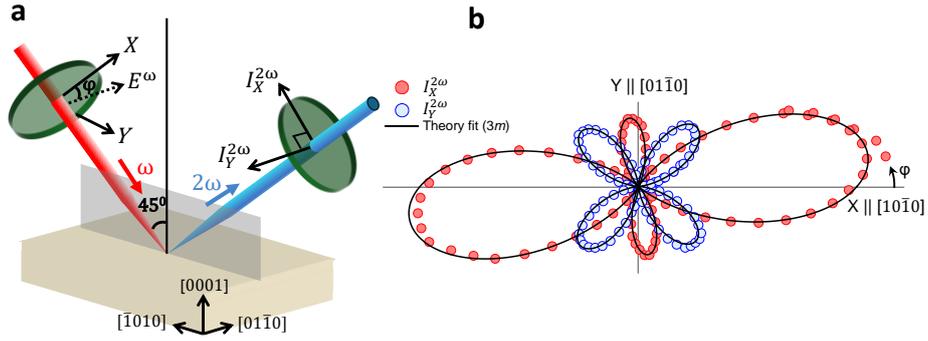

**Figure 2.** Second harmonic generation (SHG) in tr-Cr$_5$Te$_8$. (a) Schematic of the SHG setup used for measurements. An 800 nm fundamental wavelength is used to generate 400 nm second harmonic light. The angle of incidence of measurement is 45°. φ denotes the angle of polarization of the incident beam. (b) Measured SHG polar plots as a function of incident polarization angle (φ) of the fundamental beam corresponding to two orthogonal directions of SHG detection ($I_X^{2\omega}$ (red), $I_Y^{2\omega}$ (blue)). Solid lines indicate the polar plot fitting to point group 3$m$ model.

The precession images constructed from the SCXRD data collected at 300 K and 100 K are shown in Figure S2. The existing literature on tr-Cr$_5$Te$_8$ suggests that it adopts the centrosymmetric space group $P\bar{3}m1$. We therefore started the refinement of the SCXRD data by considering this model. The structure solution and refinement (performed using Superflip in Jana2006)[26,27] using the $P\bar{3}m1$ space group yielded eight independent crystallographic sites, of which four are Te sites (Te1, Te2, Te3, Te4) and remaining are Cr sites (Cr1, Cr2, Cr3, Cr4). A detailed step-by-step description of the structure refinement process is discussed in the Supporting Information. Details of structure solution and refinement using the $P\bar{3}m1$ space group are tabulated in Tables S2 & S3.

In a recent report, the structure of Cr$_{1.3}$Te$_2$ was reported to be non-centrosymmetric with space group $P3m1$.[25] To check the validity of the non-centrosymmetric space group, the symmetry of the structural model was reduced from $P\bar{3}m1$ (S1) to $P3m1$ (S2), which is also prompted due to a group-subgroup relationship between these two space groups (Supp. Figure S10). The newly generated structure parameters were then refined and compared with the original $P\bar{3}m1$ model. Due to the reduction in symmetry, a total of fourteen sites were generated from eight sites. Each tellurium site (Te1, Te2, Te3, and Te4) was split into two sites. Two of the four chromium sites (Cr2 and Cr3) were split into four sites (Cr2_1, Cr2_2, Cr3_1 and Cr3_2). As the reflection conditions (IUCr Table Vol. A) for both $P3m1$ (S2) and $P\bar{3}m1$ (S1) space groups are the same, it is difficult to determine the correct space group from the SCXRD data refinement results. Therefore, both structural models were constructed, and their respective CIF files have been deposited in the CSD database (CCDC 2343363-2343364). The refinement results are given in Tables S2 & S4.

Figure 1k shows the PXRD data fitted with $P3m1$ (S2) space group using the Rietveld refinements. The refinement parameters are given in the figure caption. A good fit can be obtained with the space group $P\bar{3}m1$ (S1) as well with almost similar fitting parameters (see Supporting Information). Therefore, as mentioned above, it remains challenging to determine conclusively whether the crystal adopts the centrosymmetric $P\bar{3}m1$ (S1) or non-centrosymmetric $P3m1$ (S2) space group based on the XRD data alone.

To address this uncertainty, we have performed second harmonic generation (SHG) experiments. Figure 2a shows the schematic of the SHG setup used. SHG is a frequency doubling non-linear optical process where two photons at the fundamental frequency ($\omega$) generate a photon of twice the frequency ($2\omega$).[28] Only non-centrosymmetric materials are SHG active, making SHG an excellent probe for detecting evidence of broken inversion symmetry. Figure 2b shows the measured SHG polar plots as a function of incident polarization of the fundamental beam from tr-Cr$_5$Te$_8$ single crystal, signifying the non-centrosymmetric structure of the crystal.

The induced second harmonic polarization ($\boldsymbol{P^{2\omega}}$) in any material depends on the incident fundamental electric field ($\boldsymbol{E^\omega}$) through the nonlinear susceptibility tensor ($d_{ijk}$) as $\boldsymbol{P_i^{2\omega}} \propto d_{ijk} \boldsymbol{E_j^\omega} \boldsymbol{E_k^\omega}$. Here, ($i,j,k$) refer to the lab coordinate system (X, Y, Z) which are tied to the crystal axes of tr-Cr$_5$Te$_8$: ($X \parallel [10\bar{1}0]$, $Y \parallel [01\bar{1}0]$, $Z \parallel [0001]$). The incident electric field, $\boldsymbol{E^\omega}$, can be written as $\boldsymbol{E_\omega} = (E_\omega \cos(\varphi), E_\omega \sin(\varphi), 0)$ ($\varphi$ is the angle of the polarization as shown in Figure 2a-2b).

For point group 3$m$, $d_{ijk}$ can be written in Voight notation as follows:[28]

$$d_{ij} = \begin{pmatrix} 0 & 0 & 0 & 0 & d_{15} & -d_{22} \\ -d_{22} & d_{22} & 0 & d_{15} & 0 & 0 \\ d_{31} & d_{31} & d_{33} & 0 & 0 & 0 \end{pmatrix}$$

From this the expected SHG intensity for two orthogonal directions (X and Y) can be calculated as follows for the case 45° oblique incidence angle of the fundamental beam:

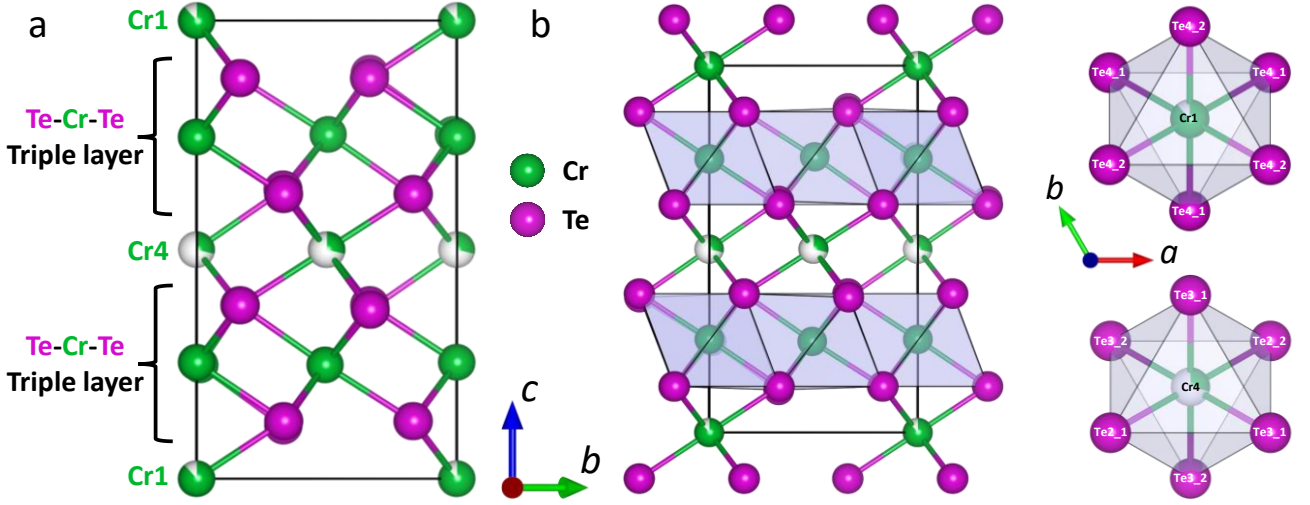

**Figure 3.** Crystal structure of tr-$Cr_5Te_8$. (a) Alternate layers of Cr and Te in the unit cell of tr-$Cr_5Te_8$. (b) Stuffed $CdI_2$-type tr-$Cr_5Te_8$.

$$I_X^{2\omega} \propto (P_X^{2\omega})^2 \propto \left((2d_{15} - d_{31} - d_{33})\cos[\varphi]^2 - 2d_{31}\sin[\varphi]^2 + \sqrt{2}d_{22}\sin[2\varphi]\right)^2$$

$$I_Y^{2\omega} \propto (P_Y^{2\omega})^2 \propto (d_{22} - 3d_{22}\cos[2\varphi] - 2\sqrt{2}d_{15}\sin[2\varphi])^2$$

The proportionality constants depend on the absolute fluence of the laser and Fresnel coefficients for the air-sample interface. These equations can be used to fit the measured polar plots confirming the observation of $3m$ symmetry in tr-$Cr_5Te_8$.

The XRD and SHG experiments together indicate that tr-$Cr_5Te_8$ crystallizes in the non-centrosymmetric space group $P3m1$ (S2) with 14 different crystallographic sites with the cell parameters $a = 7.8153(3)$ Å, $c = 11.9807(3)$ Å. The six independent crystallographic sites are occupied by Cr atoms, while the remaining eight crystallographic sites are occupied by Te atoms (Figure 3a). The structure can be viewed as a stuffed $CdI_2$-type structure (Figure 3b).

The split Cr sites (Cr2_1, Cr2_2, Cr3_1, and Cr3_2) are surrounded by six Te atoms to form $CrTe_6$ octahedra. Cr2_1 and Cr2_2 centered octahedra are linked together by edge-sharing to form a $CdI_2$-type slab. Similarly, another $CdI_2$-type slab is formed by the edge sharing of Cr3_1 and Cr3_2-centered octahedra. These slabs are alternately stacked along the [0001] direction. The other two partially occupied Cr atoms (Cr1 and Cr4) are located in octahedral interstices and alternate between two adjacent slabs.

The temperature-dependent magnetization ($M$-$T$ curve), with $B = 0.1$ T applied along the $c$-axis, shown in Figure 4a, indicates that tr-$Cr_5Te_8$ undergoes a ferromagnetic (FM) ordering with a Curie Temperature ($T_C$) ~ 200 K. A clear bifurcation is observed between the FC and ZFC curves as the temperature is lowered, indicating a significant magnetocrystalline anisotropy in tr-$Cr_5Te_8$. When the same field is applied in the $ab$-plane (Figure S6a), the magnetization value decreases by almost an order of magnitude and its variation with temperature, $M(T)$, looks more like an antiferromagnetic $M$-$T$ curve, suggesting a possible canting of magnetic spins away from the $c$-axis.

The magnetic field-dependent magnetization ($M$-$B$ curve) at temperature 2 K with the magnetic field applied along the $c$-axis in the range of $\pm 1$ T ($B_{max}=\pm 1$ T) is shown in Figure 4b. The magnetization saturates at the field $B_S \sim \pm 0.35$ T and the value of the saturation magnetization is $M_S \sim 2$ $\mu_B$/Cr, which is in agreement with the reported values.[29-31] When the field is reduced from $B_{max}=\pm 1$ T, $M(B)$ does not follow the expected trajectory near $B_S$ ($-B_S$) and remains saturated up to a critical field of $\sim 0.18$ T ($-0.18$ T). At this critical field, the magnetization undergoes a sudden, sharp drop, resembling a "hump", after which it returns to its typical path. Such a drop in magnetization is often reminiscent of a spin-flip transition, where a portion of the fully polarized spins reverses, leading to the formation of a multi-domain structure. This anomaly has been observed previously in some ferromagnetic materials, however, the underlying mechanism remains underexplored.[32-37]

To understand this anomaly further, we measured the magnetic hysteresis for various values of maximum field ($B_{max}$) and the result is plotted in Figure 4c. For $B_{max}$ values below $\pm 0.5$ T, no humps are observed on either side. It is only when $B_{max}$ reaches $\pm 0.5$ T or higher that the humps begin to appear. The hump size saturates once $B_{max}$ reaches $\pm 0.65$ T. These observations suggest that the appearance of humps depends on the magnetic history of the sample. If the $B_{max}$ visited by the sample is not large enough, the humps do not appear. A plausible explanation for these observed features is the presence of a secondary magnetic phase characterized by distinct coercivity and saturation fields compared to the primary phase. The sharp drop in magnetization may correspond to the reversal of this secondary phase. The absence of humps at lower $B_{max}$ values implies that the secondary phase begins to populate only after a certain threshold $B_{max}$ is reached (we call this $B^b_{max}$). Additionally, the saturation of hump size at higher $B_{max}$ indicates that the population of the secondary phase is fully established at this field (we call this $B^e_{max}$). In this context, the hump can be interpreted as a manifestation of magnetic memory associated with the secondary magnetic phase. Interestingly, such memory effect has very recently been observed in the ferromagnetic Weyl semimetal $Co_3Sn_2S_2$.[38] Once the secondary phase starts populating, it retains a magnetic memory. Even when the field is reduced, the secondary phase preserves its state due to its distinct coercivity, resulting in the observed hump. Figure 4d presents a *T-B* phase diagram that shows the temperature dependence of $B^b_{max}$ and $B^e_{max}$. These two critical fields divide the phase diagram into three distinct regions. In Region I, the secondary phase is not populated, and no humps are observed in the magnetisation curve. The population of the secondary magnetic phase takes place in Region II. As one moves higher within this region, the hump size progressively increases, indicating a growing population of the secondary magnetic phase. In Region III, the secondary phase is fully populated, and the hump size reaches saturation, showing no further increase with the field. This phase diagram provides a clear framework for understanding the emergence and behavior of the secondary magnetic phase across different temperatures and magnetic fields.

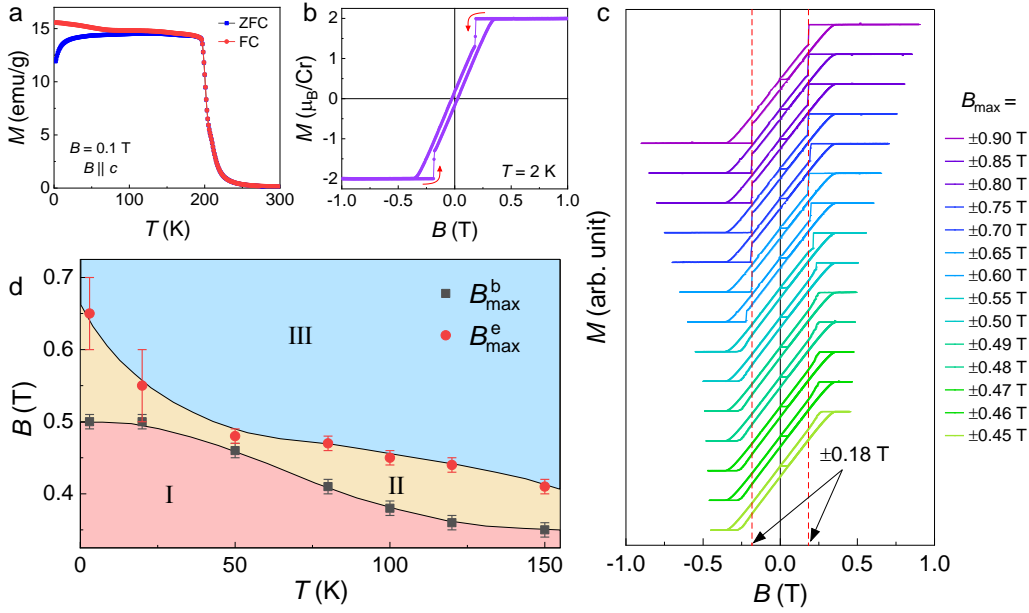

**Figure 4.** Magnetic properties of tr-$Cr_5Te_8$ with magnetic field applied along the *c*-axis. (a) Temperature dependence of the magnetization (*M*) under ZFC and FC conditions. (b)-(c) Magnetic field dependence of the magnetization at temperature 2 K (b) for field range ($B_{max}$) of ±1 T (arrows show the direction of field sweep.) (c) for various values of $B_{max}$. Red dashed lines mark the saturation of hump size. (d) *T-B* phase diagram showing the temperature variation of $B^b_{max}$ and $B^e_{max}$ and three distinct regions based on the population of secondary magnetic phase (see main text).

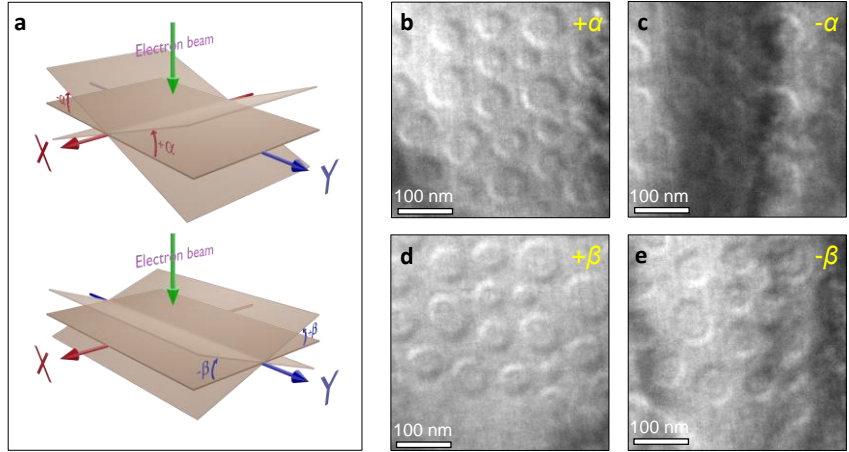

**Figure 5.** Lorentz transmission electron microscopy (LTEM) images of Néel type skyrmions in tr-$Cr_5Te_8$ at temperature 100 K in zero magnetic field. (a) Definition of tilt angles *α*, *β* and the corresponding X and Y tilt axes. LTEM images for the tilt angle of (b)-(c) *α* = ±15° and (d)-(e) *β* = ±15°.

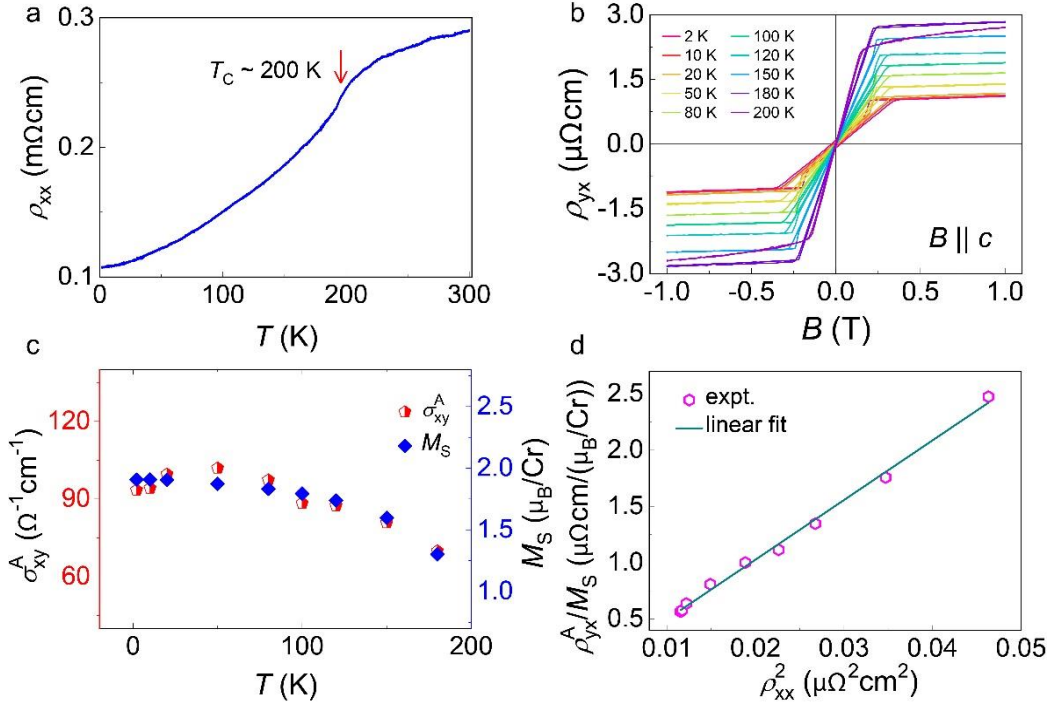

**Figure 6.** Transport properties of tr-Cr$_5$Te$_8$. (a) Temperature dependence of the longitudinal resistivity ($\rho_{xx}$). (b) Dependence of the Hall resistivity ($\rho_{yx}$) at various temperatures with the magnetic field applied along the $c$-axis. (c) Variation of anomalous Hall conductivity ($\sigma_{xy}^A$) (left axis) and saturation magnetization ($M_S$) (right axis) with temperature. (d) Scaling of the normalized anomalous Hall resistivity with the longitudinal resistivity. The green line is the linear fit of the data.

To investigate the nature of the magnetic domain in tr-Cr$_5$Te$_8$, LTEM experiments were performed on a thin $c$-axis oriented lamella of the compound. LTEM images were taken at a defocus value of 1.5 mm at zero magnetic field after cooling the sample from room temperature to 100 K in a magnetic field of 100 Oe. Sharp circular features, indicative of magnetic textures, were observed when the lamella was slightly tilted away from the electron beam during the observation. The tilt angles $\alpha$ (about X-axis), $\beta$ (about Y-axis), and the corresponding tilt axes are defined in Figure 5a. The circular features show dark and bright contrasts along their opposite edges. A reversal of the contrasts was observed for both tilt axes when the lamella was tilted in the opposite direction, i.e., the contrasts were reversed for +$\alpha$ (+$\beta$) and -$\alpha$ (-$\beta$) tilts. Figure 5b-5e shows the LTEM images taken at a tilt angle of $\pm 15°$ about the X- and Y-axes. A clear 90° contrast rotation was also observed between the α- and β-tilt images. These observations, which are key identifying features of Néel-type skyrmions,[25,39-41] suggest that the magnetic textures in tr-Cr$_5$Te$_8$ are of Néel-type. However, the skyrmions do not form a regular lattice of the underlying crystal symmetry, suggesting that in addition to the DMI, a significant long-range dipolar interaction plays an important role in stabilizing them. The skyrmions observed in tr-Cr$_5$Te$_8$ are similar to those observed in Cr$_{1.3}$Te$_2$, suggesting that these two compounds are not very different magnetically.

Figure 6a shows the temperature dependence of the longitudinal resistivity ($\rho_{xx}$-$T$ curve) from temperature 2 K to 300 K. The behavior of $\rho_{xx}$ is metallic throughout the temperature range with a residual resistivity ratio $RRR$ = $\rho_{xx}(300\,K)/\rho_{xx}(2\,K) \approx 3$. A clear anomaly corresponding to the ferromagnetic to paramagnetic phase transition is seen around 200 K, which is in good agreement with the $M$-$T$ data. Figure 6b shows the field-dependent Hall resistivity ($\rho_{yx}$-$B$ curve) at different temperatures with the magnetic field applied along the $c$-axis. In a ferromagnetic material, the Hall resistivity $\rho_{yx}$ generally consists of contributions from the ordinary Hall effect (OHE), arising from the magnetic Lorentz force acting on the moving charges and the anomalous Hall effect (AHE), arising from the extrinsic scattering processes and/or the intrinsic momentum space Berry curvature mechanism.[42] In certain materials, where a non-coplanar spin structure gives rise to a non-zero scalar spin chirality, an additional contribution, arising from the real space Berry curvature mechanism, called the topological Hall effect (THE), is expected in the Hall resistivity.[20,21] All these three contributions are expressed in the following empirical formula:

$$\rho_{yx} = \rho_{yx}^O + \rho_{yx}^A + \rho_{yx}^T$$
$$= R_0 B + \mu_0 R_S M + \rho_{yx}^T$$

where, $\rho_{yx}^O$, $\rho_{yx}^A$, $\rho_{yx}^T$ are the ordinary, anomalous, and topological Hall resistivities, respectively. $R_0$ and $R_S$ are the ordinary and anomalous Hall coefficients, respectively. The values of $R_0$ and $\rho_{yx}^A$ are given by the slope and the intercept of the linear fit of the $\rho_{xy}$-$B$ curve in the saturation region. $R_S$ can be calculated from the equation $\rho_{yx}^A = \mu_0 R_S M_S$ where $M_S$ can be taken from the $M$-$B$ curves. Figure 6c shows the variation of the anomalous Hall conductivity (AHC) $\sigma_{xy}^A \left( \frac{\rho_{yx}^A}{\rho_{xx}^2 + \rho_{yx}^2} \approx \frac{\rho_{yx}^A}{\rho_{xx}^2} \right)$ with temperature. AHC remains almost constant at low temperatures with a maximum value of 102 $\Omega^{-1}\text{cm}^{-1}$ at 50 K. This value of AHC is higher than those previously reported in $Cr_{1+d}Te_2$ systems.[30,43] The drop in the value of AHC at higher temperatures can be attributed to the similar drop observed in the temperature dependence of the saturation magnetization $M_S$. Thus, the variation of AHC with temperature is effectively negligible. This suggests that the intrinsic Berry curvature mechanism could be responsible for the AHE in tr-$Cr_5Te_8$, although previous reports have attributed the origin of the AHE to the extrinsic skew scattering mechanism.[30,43] The actual mechanism of the AHE can be determined by looking at the scaling nature of $\rho_{yx}^A$ with $\rho_{xx}$. In particular, a linear scaling of $\rho_{yx}^A$ with $\rho_{xx}^2$ suggests either the intrinsic Berry curvature or the extrinsic side jump mechanism of the AHE whereas a linear scaling of $\rho_{yx}^A$ with $\rho_{xx}$ suggests an extrinsic skew scattering mechanism of the AHE.[42] In Figure 6d the normalized anomalous Hall resistivity, $\rho_{yx}^A / M_S$, is plotted against $\rho_{xx}^2$ and the points are fitted well with a linear line, implying the intrinsic Berry curvature mechanism of AHE.

The intrinsic AHE and magnetism are closely related,[42] it is thus essential to determine the magnetic nature of tr-$Cr_5Te_8$. In Figure 7a, we display the density of states (DOS), where the black solid line represents the total DOS which is localised below -1 eV and dispersed near the Fermi energy ($E_F$). The DOS for individual atoms was also computed. For Cr, shown by the red solid line, we observed that the major contribution to the DOS comes from the up spin. The number density of Cr spin-up and spin-down electrons was also calculated, as shown by the red dashed line. To determine the magnetic moment of an individual Cr atom, we take the difference between the spin-up and spin-down electron densities and divide it by the number of Cr atoms. The magnetic moment of Cr thus obtained is 2.1 $\mu_B$/Cr. For Te, the DOS is almost symmetrical for up and down spins, indicating its non-magnetic nature. The total magnetic moment is thus contributed by Cr only, the value of which is in good agreement with the experimental value.

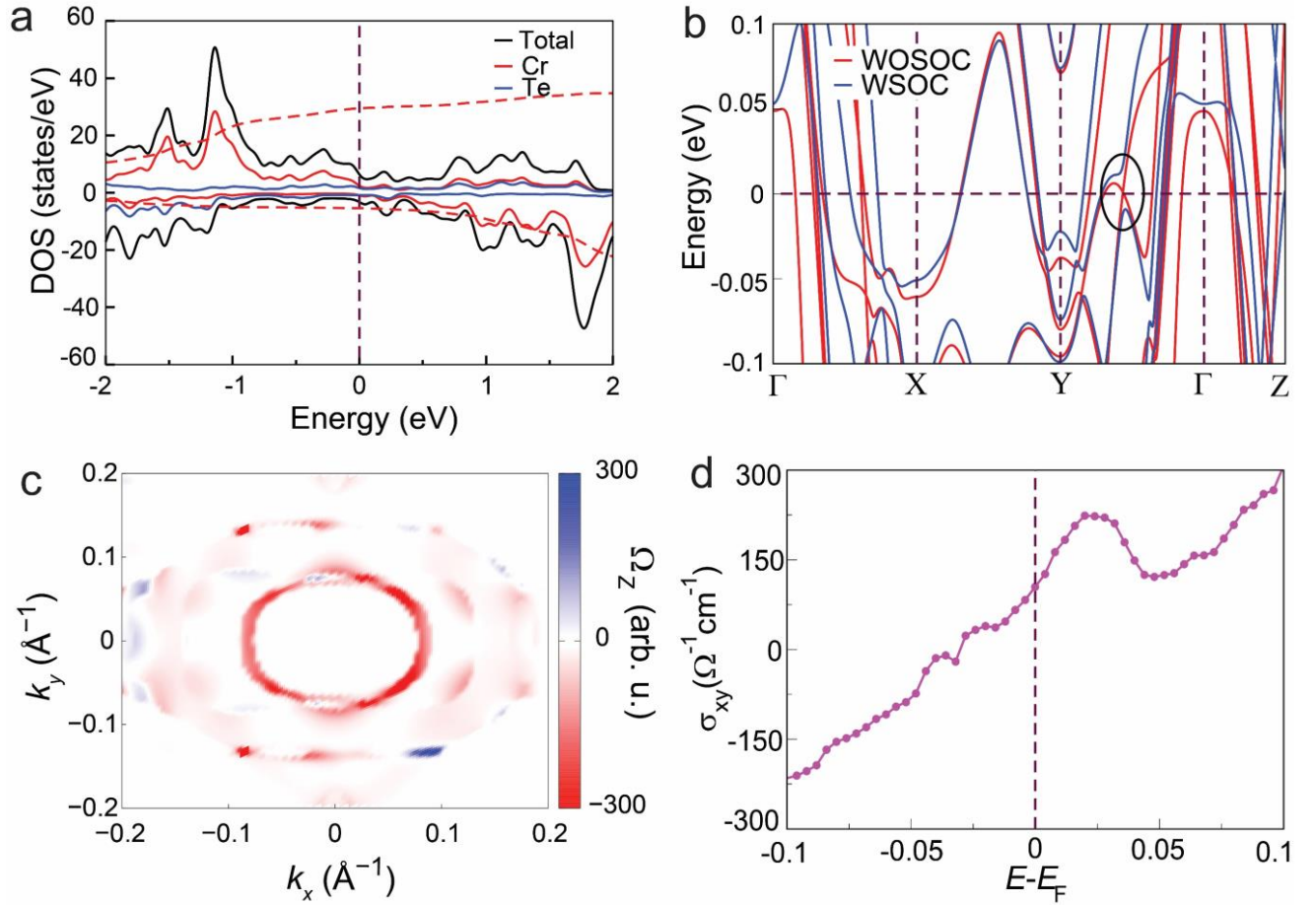

**Figure 7.** First principal calculations of tr-$Cr_5Te_8$. (a) Density of states of tr-$Cr_5Te_8$. (b) Band structure of tr-$Cr_5Te_8$ without SOC (red) and with SOC (blue). (c) Berry curvature distribution in the $k_z = 0$ plane along the gapped nodal line. (d) Energy ($E$-$E_F$) dependence of the AHC of tr-$Cr_5Te_8$.

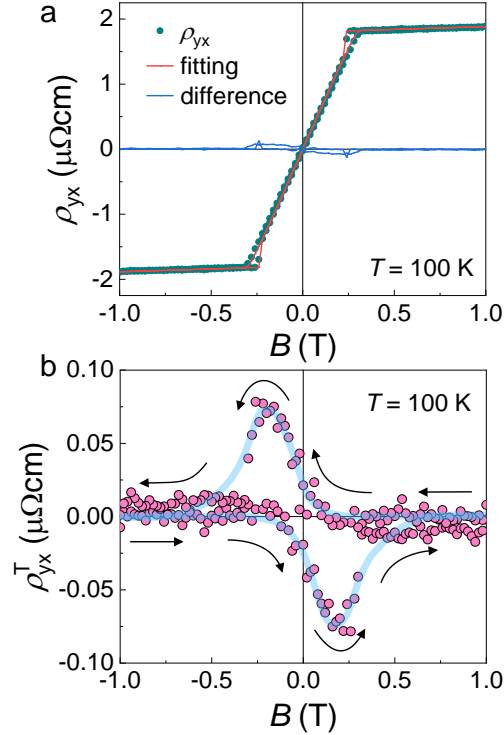

**Figure 8.** Topological Hall resistivity in tr-$Cr_5Te_8$. (a) Fitting of the total Hall resistivity by the combined contribution of ordinary and anomalous Hall resistivities and the estimation of the possible topological Hall resistivity. (b) The scatter plot of the field dependence of topological Hall resistivity at 100 K. The blue line is a guide to the eye.

In order to understand the topological features and transport properties of tr-$Cr_5Te_8$, band dispersions are determined in the absence and presence of spin-orbit coupling (SOC) (Figure 7b) using a plane-wave based pseudopotential. The band structure features a non-trivial crossing point (nodal line) near $E_F$ along the Γ-Y symmetry path in the absence of SOC. When the SOC is switched on the nodal line is gapped out (indicated by a black circle) which gives rise to a substantial Berry curvature at $k_z = 0$ (Figure 7c) plane. The Berry curvature is intrinsically linked to the AHC, as detailed in the Supporting Information. In Figure 7d, the AHC resulting from the Berry curvature mechanism is plotted as a function of the doping level. At $E_F$, the value of AHC is 104 $Ω^{-1}cm^{-1}$ which is in excellent agreement with the experimental value suggesting that the intrinsic mechanism is dominant in tr-$Cr_5Te_8$.

Skyrmions are expected to generate an additional contribution, $ρ_{yx}^T$, to the Hall resistivity through Berry curvature mechanisms.[44-46] In the adiabatic limit, the topological Hall resistivity arising from a given skyrmion density can be estimated using the expression $ρ_{yx}^T = PR_0B_{eff}$.[21,46,47] Here $P$ is the spin polarisation of the conduction electrons and $B_{eff}$ is the fictitious effective magnetic field, which is related to $n_{sk}$ via the magnetic flux quantum $Φ_0$ as $B_{eff} = n_{sk}Φ_0$. The factor $P$ can be calculated as the ratio of the saturation moment of $Cr^{3+}$ (considering that the oxidation state of Cr in tr-$Cr_5Te_8$ is $Cr^{3+}$) in tr-$Cr_5Te_8$ to the effective moment of a free $Cr^{3+}$ ion. In tr-$Cr_5Te_8$ we get $P = 0.52$, $R_0 = 1.3 \times 10^{-3}$ $cm^{-3}C^{-1}$ at 2 K and $n_{sk} \sim 1.4 \times 10^{10}$ $cm^{-2}$ which gives $B_{eff} = 2.9$ T. The estimated value of the topological Hall effect using these parameters is $\sim 1.97 \times 10^{-2}$ μΩcm. Figure 8a shows the field dependent Hall resistivity data at 100 K fitted with the combined contribution of OHE and the AHE ($ρ_{yx}^O + ρ_{yx}^A$). The difference between the experimental data and the fitted curve gives $ρ_{yx}^T$. Figure 8b shows a scatter plot of $ρ_{yx}^T$ (the difference curve of Figure 4a) with the thick blue line serving as a guide to the eye. For clarity, the kinks appearing on either side of the difference curve in Figure 8a have been removed in Figure 8b as they are just the manifestations of the inconsistency of the hump size across different measurements (see Figure S7). We observe a notable topological Hall resistivity reaching a maximum value $ρ_{yx}^{T,max} \sim 7.2 \times 10^{-2}$ μΩcm at 0.2 T. Notably, $ρ_{yx}^T$ has a non-zero value at zero magnetic field (spontaneous topological Hall resistivity, $ρ_{yx}^{T,sp}$) of magnitude $\sim 2.1 \times 10^{-2}$ μΩcm. This value of $ρ_{yx}^{T,sp}$ matches quite well with the estimated value of $ρ_{yx}^T$. This makes much sense as the zero-field skyrmion density was used to estimate $ρ_{yx}^T$.

## 2. Conclusions

We have conducted a comprehensive analysis of the crystal structure of trigonal $Cr_5Te_8$ using single crystal and powder X-ray diffraction techniques. Second harmonic generation (SHG) experiments indicate that the compound crystallizes in the non-centrosymmetric space group $P3m1$. However, a degree of ambiguity remains between the centrosymmetric $P\bar{3}m1$ and non-centrosymmetric $P3m1$ space groups based on XRD data. Given the sensitivity of the structure to the amount of Cr intercalation, the possibility of coexisting centrosymmetric and non-centrosymmetric phases cannot be ruled out, warranting further composition-dependent studies of the space group. More importantly, in support of the non-centrosymmetric structure, we have observed Néel-type skyrmions in thin lamella of tr-$Cr_5Te_8$ by Lorentz transmission electron microscopy, which are found to be stable even in the zero magnetic field. Consequently, a significant spontaneous topological Hall resistivity (~$2.1 \times 10^{-2}$ μΩcm at 100 K) originating from the skyrmion phase is observed in tr-$Cr_5Te_8$. A large anomalous Hall conductivity of 102 $\Omega^{-1}cm^{-1}$ at 50 K is observed. Scaling analysis and theoretical calculations confirm that the anomalous Hall effect originates from the intrinsic Berry curvature mechanism rather than the previously reported extrinsic skew scattering mechanism.

## 3. Experimental Section

Single crystals of tr-$Cr_5Te_8$ were grown by the self-flux method. Pieces of Cr (99.99%, *Alfa Aesar*) and Te (99.999%, *Alfa Aesar*), in the molar ratio Cr : Te = 18 : 82, were taken in a quartz crucible and sealed in a quartz tube under vacuum. The tube was then placed in a muffle furnace, heated to 1273 K, held at this temperature for 10 h, and cooled slowly (2 K/h) to 1073 K. After being held at 1073 K for 34 h, the tube was immediately removed from the furnace and centrifuged to remove the excess flux.

The elemental ratio was checked using the energy dispersive X-ray spectroscopy (EDXS) technique. EDXS data were collected on a field emission scanning electron microscope (Quanta 250 FEG) equipped with an Element silicon drift detector (SDD) with an accelerating voltage of 25 kV and an accumulation time of 60 s. Single crystal X-ray diffraction (SCXRD) intensities were collected using the BRUKER Photon II detector in the Bruker D8 Quest diffractometer equipped with Mo K$\alpha$ ($\lambda$ = 0.71073 Å) radiation. Apex 4 software was used for data acquisition and integration.[48] Precession images (Figure S2) were constructed using CrysAlis Pro software.[49] Powder X-ray diffraction (PXRD) experiments were performed in the 2$\theta$ range of 10° to 140° at room temperature using a Rigaku SmartLab diffractometer equipped with a 9 kW Cu K$\alpha$ ($\lambda$ = 1.5418 Å) X-ray source. The reflection profiles of the PXRD were refined using the pseudo-Voigt function (4 parameters were used). March-Dollase function was used to handle the preferred orientation along [0001] direction. Berar-Baldinozzi correction (4 asymmetric parameters) was implied for asymmetric correction of the powder profile. Jana2006 software was used for both PXRD and SCXRD data refinement.[26,27] The orientation of single crystals was verified by X-ray Laue diffractometer in reflected geometry and the patterns were analyzed by Orient Express software.

SHG measurements were performed using an 800 nm fundamental laser beam from a Spectra-Physics SOLSTICE ACE Ti: sapphire laser (pulse width ~ 80 fs, repetition rate of 1 kHz). A half-wave retarder plate was used to control the incident polarization of the fundamental beam. The generated second-harmonic light was first spectrally filtered, and then decomposed to two orthogonal components by an analyzer and finally detected by a photomultiplier tube. SHG polar plots were measured corresponding to these two orthogonal components of generated SHG as a function of the polarization of the incident fundamental light. All SHG measurements were done in reflection geometry with a 45° incidence angle and a beam size of ~ 40 μm on the sample. All SHG measurements were done at room temperature (295 K).

The temperature dependent Raman spectra were recorded on a cleaved bulk single crystal using LabRam HR Evolution Raman spectrometer (HORIBA France SAS-532 nm laser). X-ray photoemission spectroscopy (XPS) experiment was carried out on an as-grown single crystal using PHI 5000 Versa Prob III (Ulvac-PHI, INC).

For transmission electron microscopy (TEM) investigations, lamellae were prepared from the single crystal of tr-$Cr_5Te_8$ using the focused ion beam (FIB) Ga+ ion milling technique. This process was carried out with an FEI Nova Nanolab 600 SEM/FIB system operating at 30 keV ion-beam energy, utilizing standard lift-out procedures. Subsequent polishing of the lamellae was performed at lower Ga+ ion-beam energies (5 keV) to minimize the thickness of any amorphous surface layers. High-resolution TEM (HRTEM) imaging was conducted utilizing the JEOL ARM300F2 TEM. Magnetic textures were imaged using a JEOL JEM-F200 TEM operating in Lorentz mode at an acceleration voltage of 200 keV. A GATAN double-tilt sample holder capable of varying the temperature between 100 K and 380 K was employed for these experiments. To apply a vertical magnetic field to the lamella within the TEM column, currents were passed through the coils of the objective lens. Imaging was facilitated using a Lorentz mini-lens.

A standard six-terminal method was employed to measure both the longitudinal resistivity $\rho_{xx}$ and transverse Hall resistivity $\rho_{yx}$ simultaneously at different temperatures and magnetic fields. The current was applied in the *ab*-plane and the magnetic field along the *c*-axis of the crystal sample. Magnetic measurements were carried out using the VSM (Vibrating Sample Magnetometer) option of the Physical Properties Measurement System (PPMS, DynaCool, Quantum Design, 9T). Electrical transport measurements were also carried out in PPMS, using its ETO (Electrical Transport Option) option.

**Ab initio calculations**

The electronic band structure of tr-$Cr_5Te_8$ was theoretically analysed by ab-initio DFT calculations using the VASP package[50] with a plane-wave basis set and pseudopotentials. The cut-off energy for the plane waves was set to 600 eV and the exchange correlation potential was described by the generalized-gradient approximation (GGA+U).[51] The *k*-space integrations were performed on an 8 × 8 × 8 grid. All structures were relaxed until the forces were smaller than 0.001 eV/Å. An effective *U* of 2 eV was required to match the magnetic moments. In the next step, we derived Wannier functions from the DFT band structure using the WANNIER90 package.[52,53] The initial projections were selected as *d*-orbital for Cr and *p*-orbital for Te. Using these Wannier functions, we constructed a tight-binding Hamiltonian *H* to compute the Berry curvature in the system using the Kubo formula employing Wanniertools.[54]

**ASSOCIATED CONTENT**

**Supporting Information**

> EDXS spectrum, composition chart, fitting of Laue pattern, XRD refinement process, precession XRD images, fitting of PXRD data with P$\bar{3}$m1 space group, temperature dependent Raman spectra, magnetic property for $B \parallel ab$, magnetoresistance data, temperature dependence of $R_0$ and $R_S$, group-subgroup relation.

**Accession codes**

CCDC 2343363-2343364 contains the supplementary crystallographic information for this paper. These data can be obtained free of charge via www.ccdc.cam.ac.uk/data_request/cif, or by emailing data_request@ccdc.cam.ac.uk, or by contacting The Cambridge Crystallographic Data Centre, 12 Union Road, Cambridge CB2 1EZ, UK; fax: +44 1223336033.

**Acknowledgements**

NK acknowledges the Science and Engineering Research Board (SERB), India, for financial support through Grant Sanction No. CRG/2021/002747 and Max Planck Society for funding under Max Planck-India partner group project. SH, and VG acknowledge support from the Department of Energy Basic Sciences Division grant number DE-SC0012375 for the nonlinear optical characterization. PPJ would like to thank SERB, India (grant no. CRG/2020/004115), for financial support. SKK acknowledges the CSIR for research fellowship. This research project made use of the instrumentation facility provided by the Technical Research Centre (TRC) at the S. N. Bose National Centre for Basic Sciences, under the Department of Science and Technology, Government of India. The authors acknowledge the use of XPS experimental facilities at TCG Crest, Salt Lake Sector 5, Kolkata, and thank Dr. Ananya Banik for her assistance with the measurements.

# Supporting Information

# Peculiar magnetic and magneto-transport properties in a non-centrosymmetric self-intercalated van der Waals ferromagnet $Cr_5Te_8$

Banik Rai, Sandip Kumar Kuila, Rana Saha, Sankalpa Hazra, Chandan De, Jyotirmoy Sau, Venkatraman Gopalan, Partha Pratim Jana, Stuart S. P. Parkin, Nitesh Kumar

The elemental composition is checked using energy-dispersive X-ray spectroscopy (EDXS) on four randomely chosen single crystals and the result is tabulated in Table S1. The composition is fairly homogeneous in different regions of each single crystal as well as among all the crystals with an average composition ratio Cr : Te = 38.06 : 61.94. The chemical formula so obtained is $Cr_{4.92}Te_8$ which have been reffered to as $Cr_5Te_8$ throughout the main text for the sake of brevity and to be consistent with the literature.

Table S1. Elemental composition of few randomely chosen single crytals of tr-$Cr_5Te_8$ as verified by EDXS.

| Crystal | Region | %Cr | %Te | Crystal Average | | Total Average | | Chemical Formula |
|---|---|---|---|---|---|---|---|---|
| | | | | %Cr | %Te | %Cr | %Te | |
| 1 | a | 37.9 | 62.1 | 38.07 | 61.93 | 38.06 | 61.94 | $Cr_{4.92}Te_8$ Or $Cr_{1.23}Te_2$ |
| | b | 38.1 | 61.9 | | | | | |
| | c | 38.2 | 61.8 | | | | | |
| 2 | a | 37.9 | 62.1 | 37.90 | 62.10 | | | |
| | b | 37.7 | 62.3 | | | | | |
| | c | 38.3 | 61.7 | | | | | |
| | d | 37.7 | 62.3 | | | | | |
| 3 | a | 38.4 | 61.6 | 38.50 | 61.50 | | | |
| | b | 38.4 | 61.6 | | | | | |
| | c | 38.7 | 61.3 | | | | | |
| 4 | a | 37.4 | 62.6 | 37.77 | 62.23 | | | |
| | b | 37.9 | 62.1 | | | | | |
| | c | 38.0 | 62.0 | | | | | |

Figure S1a and S1b show the reflected Laue diffraction pattern and fitted Laue diffraction pattern, respectively. The pattern consists of sharp spots which indicates good quality of the single crystal.

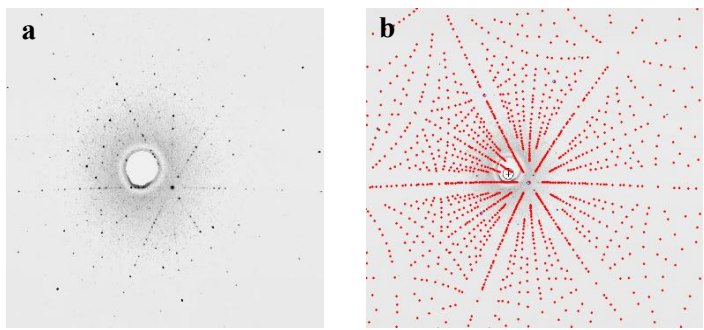

**Figure S1.** (a) Laue diffraction pattern obtained in the reflected mode along the [0001] direction of the crystal. (b) Fitted Laue pattern.

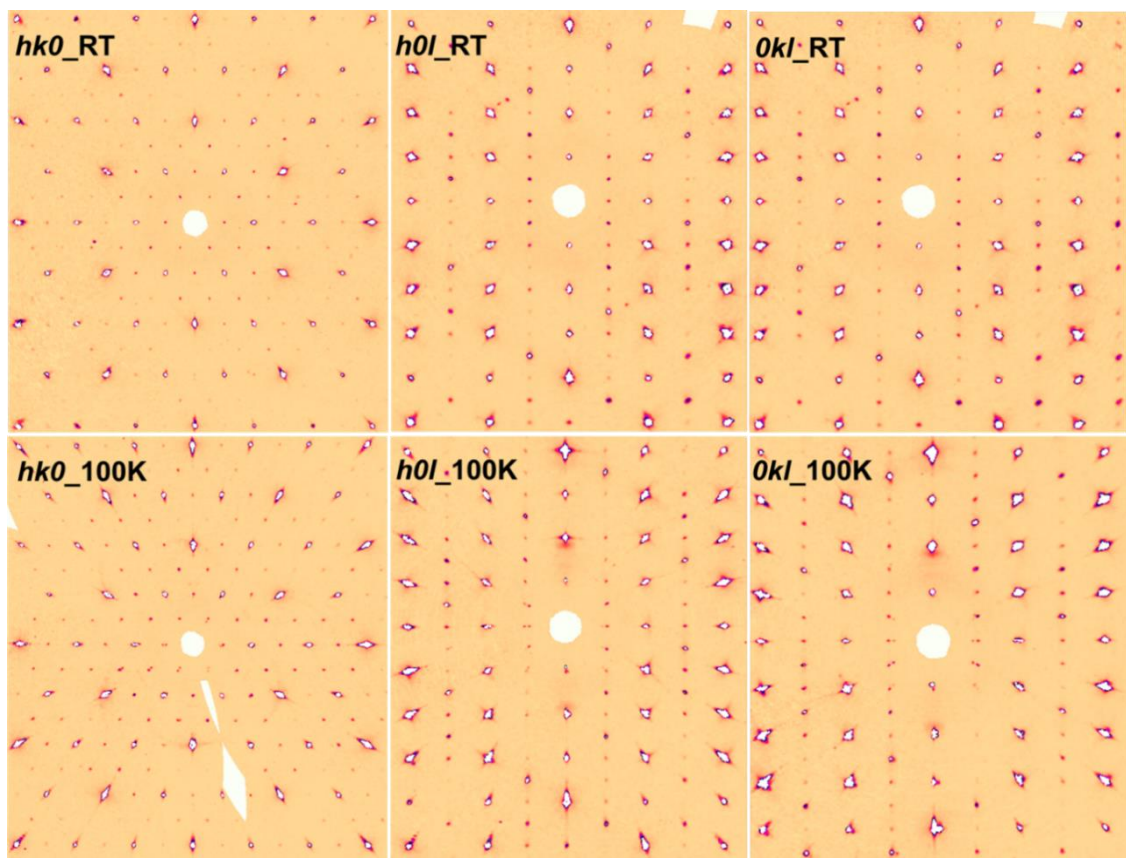

**Figure S2**. Precession unwarp images of tr-$Cr_5Te_8$ in *hk0*, *h0l*, *0kl* planes at 300 K (top) and at 100 K (bottom).

Jana 2006 was used to refine the single crystal X-ray diffraction (SCXRD) data. The space group $P\bar{3}m1$ was suggested from the symmetry search in the electron density map. At the initial stage, the structure solution obtained by Superflip yielded seven independent crystallographic sites: four Te sites (Te1, Te2, Te3, Te4) and three Cr sites (Cr1, Cr2, Cr3). Preliminary refinement resulted to $R_{obs}(F)$ = 9%. Difference Fourier map was constructed at this stage to check the missing atom if there is any. A peak maximum with large integrated charge (~4) at a reasonable distance (0 ½ ½) from Te3 (2.68 Å) was detected and included as chromium (Cr4) in the initial structural model. This Cr4 was checked for occupational disorder. Subsequent refinement converged to residual value of about 4% with SOF (Cr4) ≈ 0.30. Occupancy of other Cr sites were also refined. Refined SOF of Cr2 and Cr3 were close to one and Cr1 discerned SOF of 0.9. Hence, the SOF of Cr2 and Cr3 were changed to 1 and kept them fixed. Harmonic or anisotropic displacement parameters for all the atoms were set at this point and the refinement converged with R value about 2.5%. The final refinement including isotropic Becker & Coppens extinction correction resulted in *R-obs* ($F^2$) value of 2.5%. Refinement details using $P\bar{3}m1$ and $P3m1$ space group are tabulated in the Table S2. Structure solution using $P\bar{3}m1$ and $P3m1$ space group are tabulated in table S3 and S4, respectively.

**Table S2. Data collection and single crystal refinement details of tr-Cr$_5$Te$_8$.**

| | Cr$_5$Te$_8$ (S1) | Cr$_5$Te$_8$ (S2) |
|---|---|---|
| Nominal Composition | Cr$_5$Te$_8$ (S1) | Cr$_5$Te$_8$ (S2) |
| Chemical formula | Cr$_{4.885}$Te$_8$ | Cr$_{4.876}$Te$_8$ |
| $M_r$ | 1274.8 | 1274.3 |
| Pearson symbol | hP28 | |
| Crystal system, space group | Trigonal, $P\bar{3}m1$ | Trigonal, $P3m1$ |
| Temperature (K) | 297(2) | |
| a, c (Å) | 7.8153(3), 11.9807(3) | |
| V (Å$^3$) | 633.73(4) | |
| Z | 2 | |
| Radiation type | Mo Kα | |
| μ (mm$^{-1}$) | 22.06 | 22.06 |
| Crystal size (mm) | 0.14 × 0.11 × 0.04 | |
| Data collection | | |
| Diffractometer | Bruker Photon II | |
| Absorption correction | multi-scan | |
| $T_{min}$, $T_{max}$ | 0.419, 0.748 | |
| No. of measured, independent and observed [$I > 3\sigma(I)$] reflections | 19982, 1184, 828 | 19982, 2356, 1521 |
| $R_{int}$ | 0.054 | 0.042 |
| (sin θ/λ)$_{max}$ (Å$^{-1}$) | 0.833 | |
| Refinement | | |
| $R [F^2 > 2\sigma(F^2)]$ | 0.025 | 0.035 |
| $wR(F^2)$ (all data) | 0.105 | 0.128 |
| GOF (all data) | 1.83 | 1.70 |
| No. of reflections | 1184 | 2356 |
| No. of parameters | 37 | 67 |
| Δρ$_{max}$, Δρ$_{min}$ (e.Å$^{-3}$) | 1.79, -1.64 | 1.37, -1.57 |

**Table S3. Coordinates, site occupancy factor and isotropic displacement parameters of tr-Cr$_5$Te$_8$ ($P\bar{3}m1$, hP28, 164).**

| Atom | Wyck. | Site | S.O.F. | x/a | y/b | z/c | U [Å$^2$] |
|---|---|---|---|---|---|---|---|

| Atom | Wyck. | Site | S.O.F. | x/a | y/b | z/c | U [Å²] |
|---|---|---|---|---|---|---|---|
| Te1 | 2d | 3m. | 1 | 1/3 | 2/3 | 0.88347(4) | 0.01141(13) |
| Te2 | 2d | 3m. | 1 | 2/3 | 1/3 | 0.63090(4) | 0.01033(13) |
| Te3 | 6i | .m. | 1 | 0.16702(1) | 0.33404(2) | 0.62107(2) | 0.01109(12) |
| Te4 | 6i | .m. | 1 | 0.83273(1) | 0.16727(1) | 0.87389(2) | 0.01075(12) |
| Cr1 | 1a | -3m. | 0.891(6) | 0 | 0 | 0 | 0.0077(3) |
| Cr2 | 2c | 3m. | 1 | 0 | 0 | 0.25471(8) | 0.0115(2) |
| Cr3 | 6i | .m. | 1 | 0.50680(4) | 0.49320(4) | 0.24871(5) | 0.0113(2) |
| Cr4 | 3f | .2/m. | 0.293(4) | 1/2 | 0 | 1/2 | 0.0143(10) |

**Table S4. Coordinates, site occupancy factor and isotropic displacement parameters of tr-Cr$_5$Te$_8$ (*P*3*m*1, *hP*28, 156).**

| Atom | Wyck. | Site | S.O.F. | x/a | y/b | z/c | U [Å²] |
|---|---|---|---|---|---|---|---|
| Te1_1 | 1b | 3m. | 1 | 1/3 | 2/3 | 0.88342(10) | 0.0103(4) |
| Te1_2 | 1c | 3m. | 1 | -1/3 | -2/3 | -0.88362(12) | 0.0126(4) |
| Te2_1 | 1c | 3m. | 1 | 2/3 | 1/3 | 0.63092(12) | 0.0119(4) |
| Te2_2 | 1b | 3m. | 1 | -2/3 | -1/3 | -0.63091(12) | 0.0091(4) |
| Te3_1 | 3d | .m. | 1 | 0.16681(6) | 0.33361(11) | 0.62090(9) | 0.0115(3) |
| Te3_2 | 3d | .m. | 1 | -0.16724(6) | -0.33447(11) | -0.62130(8) | 0.0110(3) |
| Te4_1 | 3d | .m. | 1 | 0.83284(6) | 0.16716(6) | 0.87376(10) | 0.0094(2) |
| Te4_2 | 3d | .m. | 1 | -0.83263(6) | -0.16736(6) | -0.87420(9) | 0.0125(3) |
| Cr1 | 1a | 3m. | 0.883(5) | 0 | 0 | 0.0015(7) | 0.0075(3) |
| Cr2_1 | 1a | 3m. | 1 | 0 | 0 | 0.2545(4) | 0.0118(8) |
| Cr2_2 | 1a | 3m. | 1 | 0 | 0 | -0.2547(5) | 0.0126(9) |
| Cr3_1 | 3d | .m. | 1 | 0.50700(16) | 0.49300(16) | 0.2489(4) | 0.0128(6) |
| Cr3_2 | 3d | .m. | 1 | -0.50669(16) | -0.49331(16) | -0.2483(4) | 0.0108(5) |
| Cr4 | 3d | .m. | 0.289(4) | 0.5035(8) | 0.0071(16) | 0.5029(9) | 0.0115(12) |

**Table S5. Crystallographic environment of Te atoms.**

| Atom | Wyck position | Surrounding atoms | count | distance | Total no. of surrounding Cr atoms |
|---|---|---|---|---|---|
| Te1_1 | 1b | Cr3_2 | 3 | 2.6795(30) | 3 |
| Te1_2 | 1c | Cr3_1 | 3 | 2.6818(31) | 3 |
| Te2_1 | 1c | Cr4 | 3 | 2.6885(108) | 3.867 |
|  |  | Cr3_2 | 3 | 2.7569(28) |  |
| Te2_2 | 1b | Cr4 | 3 | 2.7568(28) | 3.867 |
|  |  | Cr3_1 | 3 | 2.8070(108) |  |
| Te3_1 | 3d | Cr4 | 1 | 2.6821(117) | 3.578 |
|  |  | Cr4 | 1 | 2.6826(74) |  |
|  |  | Cr2_2 | 1 | 2.7055(34) |  |
|  |  | Cr3_2 | 2 | 2.7102(30) |  |
| Te3_2 | 3d | Cr4 | 1 | 2.6793(75) | 3.578 |
|  |  | Cr4 | 1 | 2.6799(118) |  |
|  |  | Cr2_2 | 2 | 2.6991(30) |  |
|  |  | Cr3_2 | 1 | 2.7090(28) |  |
| Te4_1 | 3d | Cr3_2 | 2 | 2.7252(28) | 3.883 |
|  |  | Cr1 | 1 | 2.7317(48) |  |
|  |  | Cr2_2 | 1 | 2.7366(35) |  |
| Te4_2 | 3d | Cr1 | 1 | 2.7112(47) | 3.883 |

| | | Cr3_1 | 2 | 2.7326(28) | |
| | | Cr2_1 | 1 | 2.7405(28) | |

Table S5 summarizes the crystallographic environment of the Te sites. Based on the number of Cr atoms surrounding each Te atom, we identify three distinct types of Te. Within a unit cell, two Te atoms are coordinated by three Cr atoms, six Te atoms are coordinated by an average of 3.6 Cr atoms, and eight Te atoms are coordinated by an average of 3.9 Cr atoms. Given that X-ray Photoelectron Spectroscopy (XPS) is sensitive to the local chemical environment, this variation is likely to cause the Te peak to split into multiple components, contributing to the observed distortion in the overall peak profile.

Figure S3 shows the powder X-ray diffraction (PXRD) pattern of tr-$Cr_5Te_8$ fitted with the space group $P\bar{3}m1$ using the Rietveld refinement. The fitting parameters are: $R$(obs)= 2.42, $wR$(obs)= 4.04, $R$(all)= 2.62, $wR$(all)= 4.11; GOF= 2.24, $Rp$= 4.13, $wRp$= 5.93. The fitting parameters are comparable to those obtained when the PXRD pattern were fitted with the non-centrosymmetric space group $P3m1$.

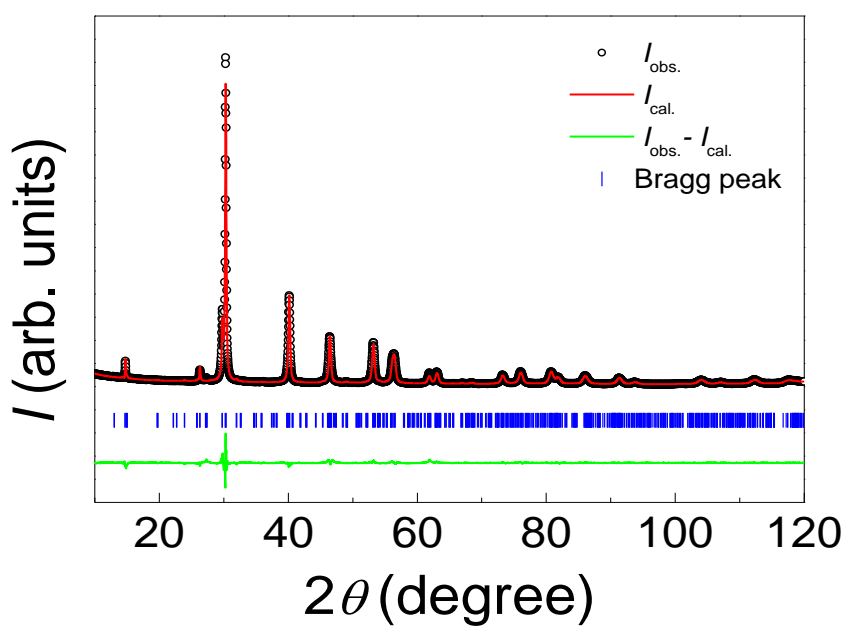

**Figure S3.** X-ray diffractogram of tr-$Cr_5Te_8$ refined with $P\bar{3}m1$ space group using the Rietveld refinement.

Figure S4 shows the temperature dependent Raman spectra of tr-$Cr_5Te_8$. We did not observe any extra peak down to 88 K confirming that there is no structural transition in tr-$Cr_5Te_8$ at low temperatures.

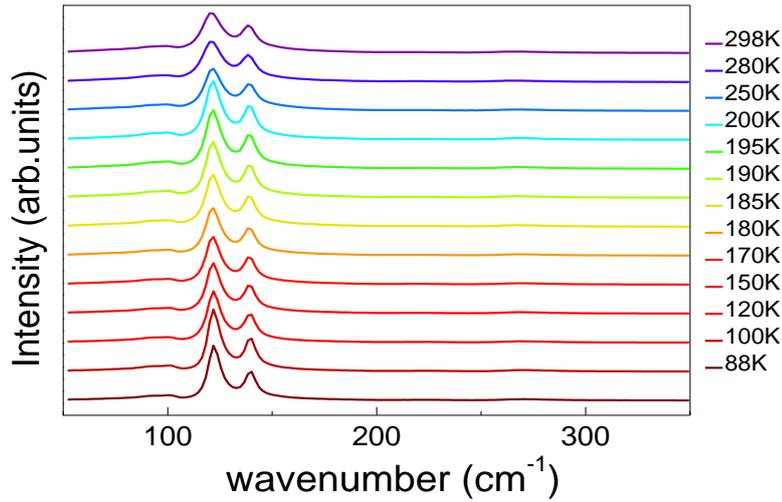

**Figure S4.** Raman spectra of tr-Cr$_5$Te$_8$ at various temperatures. Spectra obtained at different temperatures are shifted vertically for clarity.

Figure S5 illustrates the magnetic field dependent isothermal magnetisation of tr-Cr$_5$Te$_8$ at various temperatures throughout the FM phase with magnetic field applied along the *c*-axis. Although the size of the hump is found to be decreasing with increasing temperature, its appearance only on the positive side of the curve is consistent for the field range of ±0.5 T [Figure S5a]. For the field range of ±1 T the hump appears on both sides of the curve but with diminishing size at higher temperatures [Figure S5b].

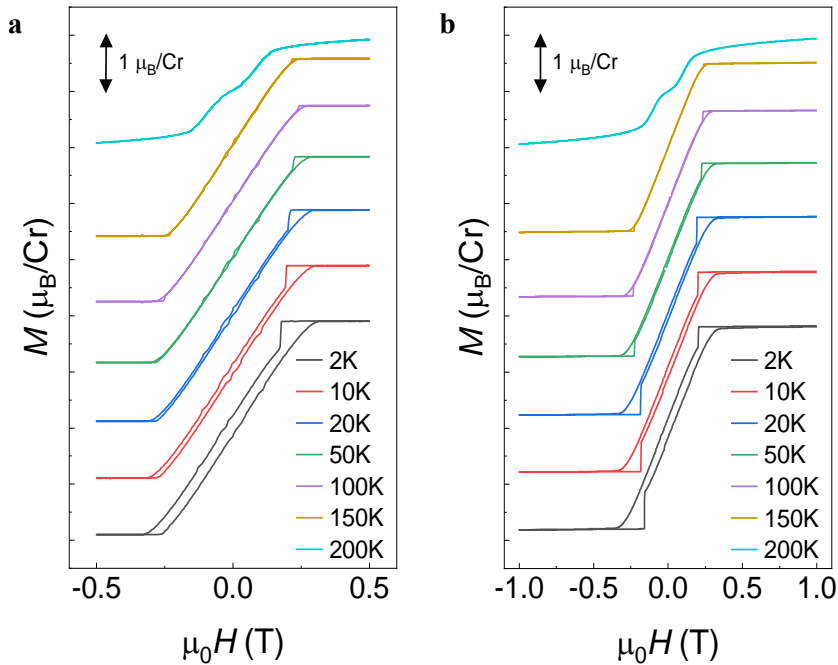

**Figure S5.** Magnetic hysteresis of tr-Cr5Te8 at various temperatures in the field range of (a) ±0.5 T and (b) ±1 T applied along the *c*-axis.

The magnetic behaviour of tr-Cr$_5$Te$_8$ when the external magnetic field is applied along the hard *ab*-plane is depicted in Fig. S6. The temperature dependent magnetization [Fig. S6(a)] shows a peak around 200K with a clear bifurcation between ZFC and FC curves at low temperatures. The curve looks like a typical antiferromagnetic magnetization curve with a Néel temperature of 200K. Because of large magnetocrystalline anisotropy, the magnetization does not saturate even at 9 T external in plane magnetic field [Fig. S6(b)].

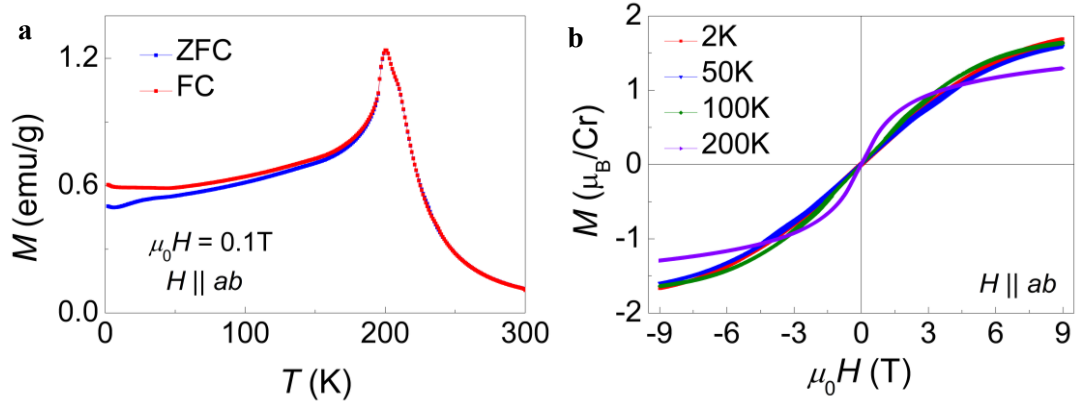

**Figure S6.** Magnetic properties of tr-$Cr_5Te_8$ with magnetic field applied along the *ab*-plane. (a) Temperature dependence of the magnetization under ZFC and FC modes in the external magnetic field of 0.1 T. (b) Magnetic field dependence of the magnetisation at various temperatures.

Figure S7 shows the repeated cycles of magnetic hysteresis at 2 K with magnetic field applied along the *c*-axis. The size of the hump, which appears only on the positive side of the curve, is not consistent in each cycle. This inconsistency arises in the Hall resistivity also which gives rise to the appearance of peaks in Figure 9 of the main text.

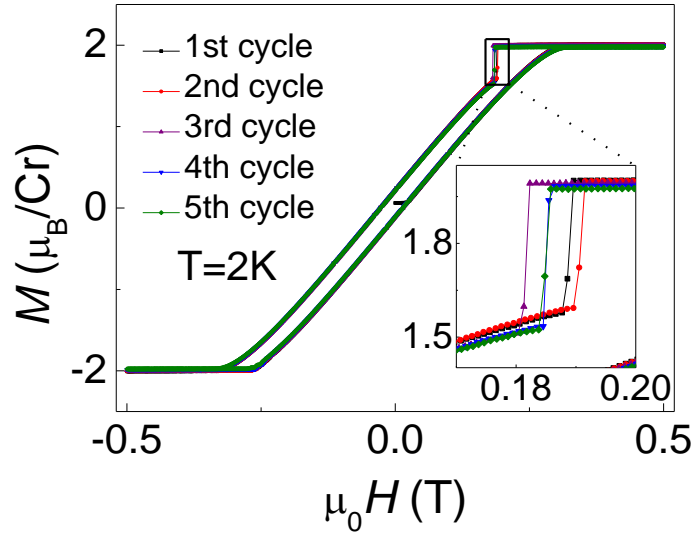

**Figure S7.** Cycles of magnetic hysteresis showing the inconsistency in the size of the humps in each cycle.

Figure S8 illustrates the magnetoresistance $\left[MR\ (\%) = \frac{\rho_{xx}(\mu_0 H) - \rho_{xx}(0)}{\rho_{xx}(0)} \times 100\right]$ of tr-$Cr_5Te_8$ with magnetic field applied along the *c*-axis. The *MR* is negative and fairly small (~ −8 % at 190 K at 9T) at almost all temperatures. The negative MR can be attributed to the suppressed scattering of conduction electrons due to spin polarization.

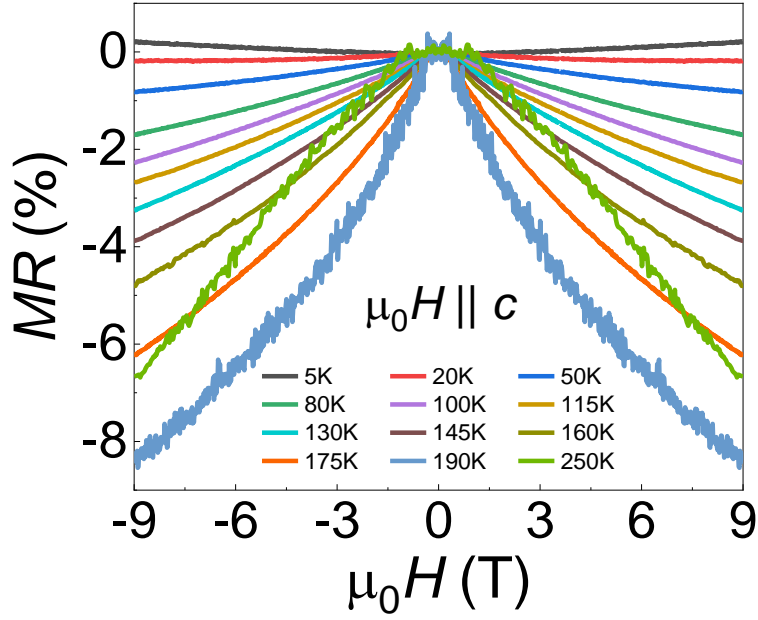

**Figure S8.** Magnetoresistance of tr-$Cr_5Te_8$ at various temperatures with magnetic field applied along the *c*-axis.

Figure S9 shows the temperature dependence of the ordinary ($R_0$) and anomalous ($R_S$) Hall coefficients. $R_0$ is positive in the entire temperature range suggesting hole type conduction in tr-$Cr_5Te_8$. The carrier density is plotted as a function of temperature in Figure S9a. $R_S$ increases monotonically with temperature [Figure S9b].

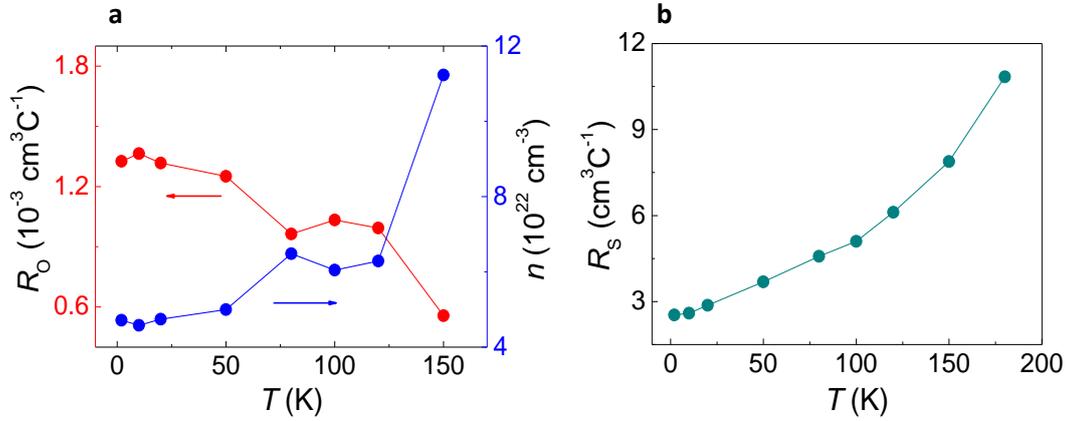

**Figure S9.** (a) Variation of ordinary Hall coefficient $R_0$ and carrier density $n$ with temperature. (b) Variation of anomalous Hall coefficient $R_S$ with temperature.

The Kubo formalism's linear response theory can be used to evaluate the intrinsic AHC, and the expression for AHC in the xy plane is

$$\sigma_{xy} = -\frac{e^2}{\hbar} \int \frac{d^3k}{(2\pi)^3} \sum_n \Omega_n^z(k) f_n(k)$$

where $\Omega_n^z$ is the Berry curvature and can be written as

$$\Omega_n^z(k) = -2i \sum_{m \neq n} \frac{\langle \psi_{nk}|v_x|\psi_{mk}\rangle \langle \psi_{mk}|v_y|\psi_{nk}\rangle - (x \leftrightarrow y)}{[E_m(k) - E_n(k)]^2}$$

where $f_n(k)$ represents the Fermi-Dirac distribution function, $n$ is the occupied band index, $E_n(k)$ is the eigenvalue of the $n^{th}$ eigenstate $\psi_{nk}$, $v_i = \frac{1}{\hbar}\frac{\partial H(k)}{\partial k_i}$ ($i = x, y, z$) is the velocity operator.

To calculate $\sigma_{xy}$, the spin-orbit coupling was considered along the [0001] direction, which is also the direction of the magnetic polarisation. Nearly degenerate bands near the nodal line in the $k_z = 0$ plane have a large Berry curvature as shown in the main text Figure 8b and contribute to the intrinsic AHC.

$P\bar{3}m1$ (no.164)

| | 2d (Te1) 3m. | 2d (Te2) 3m. | 6i (Te3) .m. | 6i (Te4) .m. | 1a (Cr1) $\bar{3}m$. | 2c (Cr2) 3m. | 6i (Cr3) .m. | 3f (Cr4) .2/m. |
|---|---|---|---|---|---|---|---|---|
| | 1/3 | 2/3 | 0.16702(1) | 0.83273(1) | 0 | 0 | 0.50680(4) | 1/2 |
| | 2/3 | 1/3 | 0.33404(2) | 0.16727(1) | 0 | 0 | 0.49320(4) | 0 |
| | 0.88347(4) | 0.63090(4) | 0.62107(2) | 0.87389(2) | 0 | 0.25471(8) | 0.24871(5) | 1/2 |

↓

$P3m1$ (no.156)

| 1b (Te1_1) 3m. | 1c (Te1_2) 3m. | 1c (Te2_1) 3m. | 1b (Te2_2) 3m. | 3d (Te3_1) .m. | 3d (Te3_2) .m. | 3d (Te4_1) .m. | 3d (Te4_2) .m. | 1a (Cr1) 3m. | 1a (Cr2_1) 3m. | 1a (Cr2_1) 3m. | 3d (Cr3_1) .m. | 3d (Cr3_1) .m. | 3d (Cr4) .m. |
|---|---|---|---|---|---|---|---|---|---|---|---|---|---|
| 1/3 | -1/3 | 2/3 | -2/3 | 0.16681(6) | -0.16724(6) | 0.83284(6) | -0.83263(6) | 0 | 0 | 0 | 0.50700(16) | -0.50669(16) | 0.5035(8) |
| 2/3 | -2/3 | 1/3 | -1/3 | 0.33361(11) | -0.33447(11) | 0.16716(6) | -0.16736(6) | 0 | 0 | 0 | 0.49300(16) | -0.49331(16) | 0.0071(16) |
| 0.88342(10) | -0.88362(12) | 0.63092(12) | -0.63091(12) | 0.62090(9) | -0.62130(8) | 0.87376(10) | -0.87420(9) | 0.0015(7) | 0.2545(4) | -0.2547(5) | 0.2489(4) | -0.2483(4) | 0.5029(9) |

**Figure S10.** Group-subgroup relationship for centro- and non-centrosymmetric space group.